\documentclass[%
 reprint,
 amsmath,amssymb,
 aps,
pra,
]{revtex4-2}

\usepackage{graphicx}
\usepackage{dcolumn}
\usepackage{bm}
\usepackage{float}

\begin{document}

\preprint{APS/123-QED}

\title{Single-photon-triggered spin squeezing with decoherence reduction in optomechanics via phase matching}

\author{Zhucheng Zhang,$^{1}$ Lei Shao,$^1$ Wangjun Lu,$^{1,6}$ Yuguo Su,$^{2}$ Yi-Ping Wang,$^{3}$ Jing Liu,$^{4}$ and Xiaoguang Wang$^{1,5}$}
\email{xgwang1208@zju.edu.cn}
\affiliation{$^{1}$Zhejiang Institute of Modern Physics, Department of Physics, Zhejiang University, Hangzhou 310027, China\\
$^{2}$State Key Laboratory of Magnetic Resonance and Atomic and Molecular Physics,\\ Wuhan Institute of Physics and Mathematics, 
APM, Chinese Academy of Sciences, Wuhan 430071, China\\
$^{3}$College of Science, Northwest A$\&$F University, Yangling 712100, China\\
$^{4}$MOE Key Laboratory of Fundamental Physical Quantities Measurement, Hubei Key Laboratory of Gravitation and Quantum Physics,\\
PGMF and School of Physics, Huazhong University of Science and Technology, Wuhan 430074, China\\
$^{5}$Graduate School of China Academy of Engineering Physics, Beijing 100193, China\\
$^{6}$Department of Maths and Physics, Hunan Institute of Engineering, Xiangtan 411104, China}
\date{\today}

\begin{abstract}
Quantum spin squeezing is an important resource for quantum information processing, but its squeezing degree is not easy to preserve in an open system with decoherence. Here, we propose a scheme to implement single-photon-triggered spin squeezing with decoherence reduction in an open system. In our system, a Dicke model (DM) with an added quadratic optomechanical coupling is considered, which can be equivalent to an effective DM manipulated by the photon number. Besides, the phonon mode of the optomechanical coupling is coupled to a squeezed-vacuum reservoir with a phase matching, resulting in that its thermal noise can be suppressed completely. We show that squeezing of the phonon mode triggered by a single photon can be transferred to the spin ensemble totally, and pairwise entanglement of the spin ensemble can be realized if and only if there is spin squeezing. Importantly, when considering the impact of the environment, our system can obtain a better squeezing degree than the optimal squeezing that can be achieved in the traditional DM. Meanwhile, the spin squeezing generated in our system is immune to the thermal noise. This work offers an effective way to generate spin squeezing with a single photon and to reduce decoherence in an open system, which will have promising applications in quantum information processing.
\end{abstract}

\maketitle

\section{Introduction}\label{sec:1}
The spin squeezed states are states with reduced quantum fluctuations in one of the collective spin components, which arises from quantum correlation between spins~\cite{1,2,3,4}.
Due to the reduced quantum fluctuations, the spin squeezed states have wide applications in the field of precision measurements, e.g., Ramsey spectroscopy~\cite{1,3,5}, atomic clocks~\cite{6,7,8}
and gravitational wave interferometers~\cite{9,10,11}. Besides, the spin squeezing can also be used as entanglement witnesses to detect quantum entanglement~\cite{12,13,14}.
Therefore, many efforts have been made to produce the squeezed states of spin systems. Specifically, the methods for generating spin squeezing are mainly divided into two categories: atom-photon interactions~\cite{3,GMPRA1989,15,16}, and nonlinear atom-atom interactions~\cite{XWPRA2003,17,19,20,21,22,23,SYBPRL,WQNP2020,QSPRA2021}. The most popular criterions of spin squeezing were proposed by Kitagawa and Ueda~\cite{2}, and by Wineland et al.~\cite{3}, i.e.,
for a system with $N$ spins represented by the collective operators $J_{\alpha}$ ($\alpha \in {x,y,z}$), the spin squeezing parameters are
\begin{equation}
\xi^2_s = \frac{4 \textrm{min} \left(\Delta J_{\vec{n}_{\perp}}\right)^{2}}{N},
\end{equation}
and
\begin{equation}
\xi_{R}^{2}=\frac{N \textrm{min} \left(\Delta J_{\vec{n}_{\perp}}\right)^{2}}{\left| \left\langle \vec{J}\right\rangle \right|^{2}},
\end{equation}
respectively, where $\vec{n}_{\perp}$ represents an axis perpendicular to the mean spin $\left\langle \vec{J}\right\rangle$,
and the minimum of variance $(\Delta J)^2$ can be obtained. $\xi^2_s<1$ or $\xi^2_R<1$ indicates that the spin system is in a squeezed state. Due to the decoherence caused by the interaction between system and environment, the squeezing of spin systems is not easy to preserve. Thus, improving the degree of squeezing and reducing the decoherence in an open system are the focus of many researches.

Cavity optomechanics is a hot research field that explores the interaction between electromagnetic field and mechanical motion, involving optical and microwave domains~\cite{24}. With this powerful platform, many theoretical and experimental advances have been made, such as optomechanically induced transparency~\cite{30GS,31SW,32AK,33MH}, ground-state cooling of mechanical
oscillators~\cite{34LQ,35CW,36YC,37JH}, quantum entanglement~\cite{38DV,39SB,40JQ,ZCOE2020}, precision measurements~\cite{41ZP,42ZZ,43TG,44WZ,45ZZ}, squeezing effects of optical and mechanical modes~\cite{25CF,26SM,27DW,28ZC,29CH,XYLPRL2015} and applications in creating phonon laser~\cite{YLZ2018NJP}, quantum states~\cite{WQPRL2021,YHCPRL2021}, and ultraprecision force sensing~\cite{44WZ}, and so on. In particular, to enhance atom-phonon coupling, the linear optomechanical interaction combined with cavity QED was investigated~\cite{46}. Besides, the quadratic optomechanical coupling was also introduced into cavity QED to explore the superradiant quantum phase transition~\cite{47,48} and the statistical properties of phonon~\cite{49}. Cavity optomechanics combined with cavity QED will obtain more abundant physical phenomena and substantially advance the fields of cavity optomechanics.

In this paper, we propose a scheme for enhancing spin squeezing and reducing decoherence in a hybrid quantum model including optomechanics and cavity QED. In our system, a Dicke model (DM) is introduced into the quadratic optomechanics, which can be equivalent to an effective DM manipulated by the photon number. Besides, the phonon mode of the optomechanical system is coupled to a squeezed-vacuum reservoir with a phase matching, resulting in that the thermal noise caused by the interaction between phonon mode and environment can be suppressed completely. Due to the quadratic optomechanical coupling, we show that the squeezing of the phonon mode triggered by a single photon can be transferred to the spin ensemble totally, and pairwise entanglement of the spin ensemble can be realized if and only if there is spin squeezing. Importantly, compared with the spin squeezing transferred from the bosonic field in the traditional DM, our system can achieve a better squeezing degree and be immune to the thermal noise, which is important and valuable to enhance the squeezing degree and reduce the decoherence in an open system.

This paper is organized as follows: In Sec.~\ref{sec:2}, we first introduce the hybrid quantum model and analyze its experimental feasibility, then derive its Hamiltonian. In Sec.~\ref{sec:3}, we give the expressions of the optimal squeezing parameters for spin ensemble and phonon mode, then analyze the evolution of the squeezing parameters with time in an open system, as well as compare the spin squeezing in our system with the one in a traditional DM. In Sec.~\ref{sec:4}, we verify some approximations used in our derivations and discuss the physical realization for our scheme. Finally, we summarize our conclusions in Sec.~\ref{sec:5}.

\section{Model and Hamiltonian}\label{sec:2}
 \begin{figure}
\centering
\includegraphics[width=1\linewidth]{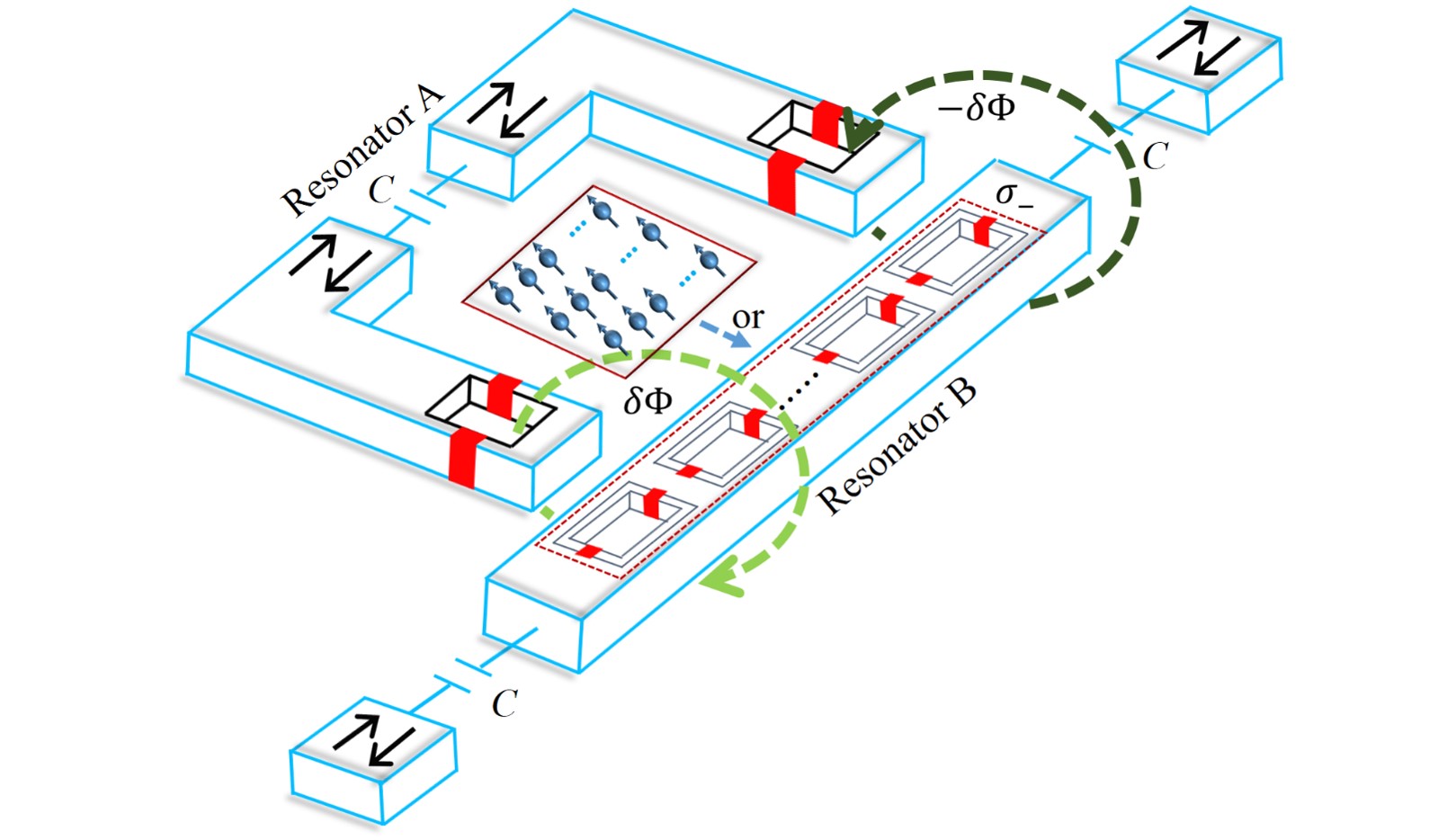}
\caption{(Color online) Schematic diagram of the system. A Dicke model [i.e., $N$ two-level systems $\sigma_-$ (e.g., a spin ensemble~\cite{50} or superconducting qubits~\cite{51}) coupled with resonator B] is introduced into a superconducting circuit that has the ability to simulate the quadratic optomechanical coupling~\cite{47,52}. In
the superconducting circuit, the photon and phonon modes are provided by resonator A and resonator B, respectively, and the antisymmetric current distribution in the resonator B ensures the opposite flux variations $\pm \delta \Phi$. By optimizing the coupling capacitance $C$, the bias flux and the geometrical arrangement of the circuit, the circuit analog of quadratic optomechanics can achieve a stronger coupling compared with the version in the membrane-in-the-middle cavity optomechanical system~\cite{MBPRA2008}.}
\label{fig:1}
\end{figure}

In our system, the quantum model considered here is a cavity-QED model with an added quadratic optomechanical coupling, as shown in Fig.~\ref{fig:1}. Specifically, a normal DM is introduced into the optomechanical system, where a spin ensemble or superconducting qubits are coupled with a resonator to form DM~\cite{50,51}, and the optomechanical system is implemented by a superconducting circuit that has the ability to simulate the quadratic optomechanical coupling~\cite{52}. By optimizing the coupling capacitance, the bias flux and the geometrical arrangement of the circuit, it can realize a stronger coupling strength compared with the one achieved in membrane-in-the-middle cavity optomechanical systems~\cite{52,MBPRA2008}. The Hamiltonian of system can be written as two parts ($\hbar=1$),
\begin{equation}
H=H_{\textrm{om}}+H_{\textrm{dm}},
\end{equation}
with
\begin{subequations}
\begin{align}
\label{eq:4a}
H_{\textrm{om}}&=\omega_{a}a^{\dagger}a-ga^{\dagger}a(b+b^{\dagger})^{2},\\
H_{\textrm{dm}}&=\Omega J_{z}+\omega_{b}b^{\dagger}b+G(b+b^{\dagger})J_{x},
\end{align}
\end{subequations}
where $H_{\textrm{om}}$ describes the Hamiltonian for the cavity field and the quadratic optomechanical coupling with $a\,(a^\dag)$ and $b\,(b^\dag)$ representing the annihilation (creation) operators of the photon and phonon modes, respectively. $\omega_a$ and $g$ are the resonance frequency of the cavity field and the quadratic optomechanical coupling strength. $H_{\textrm{dm}}$ is the Hamiltonian of DM with $\Omega$ and $\omega_b$ being the transition frequency of the spin ensemble and the resonance frequency of the phonon mode. $G$ is the collective coupling strength between the spin ensemble and the phonon mode.

The term of quadratic optomechanical coupling in \linebreak[4] Eq.~(\ref{eq:4a}) can be diagonalized with a squeezing transformation, then an effective
Hamiltonian can be derived by transforming $H$ with the squeeze operator $S(\zeta)=\exp\left[\left(\zeta^{*}b^{2}-\zeta b^{\dagger2}\right)/2\right]$, taking a fixed photon number $n$ for the cavity field, and dropping constant terms (see Appendix \ref{sec:level1} for details), i.e.,
\begin{equation}\label{eq:5}
H_{\text {eff }}=\Omega J_{z}+\omega_{n} b^{\dagger} b+G_{n}\left(b+b^{\dagger}\right) J_{x},
\end{equation}
where the squeeze parameter $\zeta=r_{n}e^{i\theta}$ with squeeze amplitude $r_{n}=\text{(}-1/4)\ln\left[1-4ng/\omega_{b}\right]$ and squeeze angle $\theta=\pi$. $\omega_{n}=\exp\left(-2r_{n}\right)\omega_{b}$ and $G_{n}=\exp\left(r_{n}\right)G$ are the transformed mechanical frequency and the coupling between spin ensemble and phonon mode. In the above squeezing transformation, the photon mode is assumed to be prepared into the Fock state $\left|n\right\rangle _{a}$, resulting in that the number operator $a^{\dagger}a$ can be seen as an algebraic number $n$. One can find that the effective Hamiltonian is just DM but manipulated by photons. Compared with the original DM in Ref.~\cite{51} and others, the coupling strength $G_n$ in our effective DM can be exponentially enhanced by the number of photons. The enhanced coupling strength is very important for the open system. Under the rotating-wave approximation (RWA) with the condition:  $\omega_{n}\approx\text{\ensuremath{\Omega}}$ and $\omega_{n}\gg G_{n}$, the effective Hamiltonian can be reduced as
\begin{equation}\label{eq:6}
H_{\textrm{eff}}=\Omega J_{z}+\omega_{n}b^{\dagger}b+\frac{G_{n}}{2}(bJ_{+}+b^{\dagger}J_{-}),
\end{equation}
which is a Tavis-Cummings model manipulated by photons, and $J_{\pm}$ are raising and lowering spin operators. Please note that the condition of RWA implies that $\exp(3r_n) \ll \omega_b / G$, so $G_n/G \ll \sqrt[3]{\omega_{b}/G}$. In our following numerical simulation, the parameters we take satisfy the condition of RWA.

\section{Squeezing parameters and their evolutions}\label{sec:3}

We assume that the spin ensemble and the phonon mode are both prepared into their lower energy state at $t=0$, then the initial state of the system is
\begin{equation}\label{eq:7}
|\psi(0)\rangle=\left(S(\zeta)|0\rangle_{b}\right) \otimes|J,-J\rangle,
\end{equation}
where $S(\zeta)\left|0\right\rangle_b$ is just a squeezed vacuum state of the phonon mode when the photon number is not zero, and in the Fock state, it can be expanded as $\sum_{k}c_{k}\left|k\right\rangle$ with~\cite{53}
\begin{subequations}
\begin{align}
&c_{0}=\frac{1}{\sqrt{\cosh \left(r_{n}\right)}}, \\
&c_{k}=\frac{(-1)^{k / 2}}{\sqrt{\cosh \left(r_{n}\right)}}\left[\frac{1}{2} e^{i \theta} \tanh (r_n)\right]^{k / 2} \frac{\sqrt{k !}}{(k / 2) !} \text { for } k \text { even}, \\
&c_{k}=0 \text { for } k \text { odd}.
\end{align}
\end{subequations}
Due to the fact that the phonon mode $b$ only contains even expansion coefficients in its initial state, it is natural to realize that we have the following expectation values under our system Hamiltonian,
\begin{align}
\langle b\rangle &=\left\langle b^{\dagger}\right\rangle=0, \\
\left\langle J_{-}\right\rangle &=\left\langle J_{+}\right\rangle=0.
\end{align}
Then, the squeezing parameters with optimal squeezing for the phonon mode and the spin ensemble can be derived as (see Appendix \ref{sec:level2} for details)~\cite{4,XWPRA2003}
\begin{align}
\label{eq:11}
&\xi_{b}^{2}=1+2\left[\left\langle b^{\dagger} b\right\rangle-\left|\left\langle b^{2}\right\rangle\right|\right], \\
\label{eq:12}
&\xi_{s}^{2}=\frac{1}{N}\!\left[\!\left\langle J_{+} J_{-}\!+\!J_{-} J_{+}\right\rangle \!-\! \sqrt{\left\langle J_{+}^{2}\!+\!J_{-}^{2}\right\rangle^{2}\!-\!\left\langle J_{+}^{2}\!-\!J_{-}^{2}\right\rangle^{2}}\right],
\end{align}
respectively.

In our system, the phonon mode is assumed to be coupled to a squeezed-vacuum reservoir with a squeezing amplitude $r_{e}$ and a reference phase $\phi_{e}$; and for the spins, the spontaneous emission is considered. Considering the system-bath interaction, the system dynamics can be described by the following master equation~\cite{53},
\begin{align} \label{eq:13}
\dot{\rho_s}=&\,i\left[\rho_s, \Omega J_{z}+\omega_{n} b^{\dagger} b+\frac{G_{n}}{2}\left(b J_{+}+b^{\dagger} J_{-}\right)\right] \notag\\
&+\frac{\kappa}{2}\left[2 J_{-} \rho_s J_{+}-J_{+} J_{-} \rho_s-\rho_s J_{+} J_{-}\right] \notag\\
&+\frac{\gamma}{2} N_{s}\left[2 b^{\dagger} \rho_s b-b b^{\dagger} \rho_s-\rho_s b b^{\dagger}\right] \notag\\
&+\frac{\gamma}{2}\left(N_{s}+1\right)\left[2 b \rho_s b^{\dagger}-b^{\dagger} b \rho_s-\rho_s b^{\dagger} b\right] \notag\\
&-\frac{\gamma}{2} M_{s}^{*}\left[2 b^{\dagger} \rho_s b^{\dagger}-b^{\dagger} b^{\dagger} \rho_s-\rho_s b^{\dagger} b^{\dagger}\right] \notag \\
&-\frac{\gamma}{2} M_{s}\left[2 b \rho_s b-b b \rho_s-\rho_s b b\right],
\end{align}
with
\begin{subequations}
\begin{align}\label{eq:14a}
N_{s}=&\,\sinh^{2}(r_{e})\cosh^{2}(r_{n})+\cosh^{2}(r_{e})\sinh^{2}(r_{n})\notag\\
&+\frac{1}{2}\sinh(2r_{n})\sinh(2r_{e})\cos(\phi_{e}),\\
\label{eq:14b}
M_{s}=&\,\left\{ \sinh(r_{n})\cosh(r_{e})\!+\!e^{-i\phi_{e}}\cosh(r_{n})\sinh(r_{e})\right\} \notag\\
&\times\!\left\{ \cosh(r_{n})\cosh(r_{e})\!+\!e^{i\phi_{e}}\sinh(r_{n})\sinh(r_{e})\right\},
\end{align}
\end{subequations}
being the effective thermal noise and the two-phonon correlation strength, respectively. $\rho_s=S^{\dagger}(\zeta)\rho S(\zeta)$ is the density operator of the system under the squeezing transformation. $\kappa$ and $\gamma$ are the decay rates of the spin ensemble and the phonon mode, respectively. In principle, the squeezed-vacuum reservoir can be generated by introducing an ancillary cavity mode that is linearly coupled to the phonon mode and driven by a broadband-squeezed vacuum field~\cite{54,55,56,TSYPRA}. From Eqs.~(\ref{eq:14a}) and (\ref{eq:14b}), one can find that there is a phase matching relationship between the squeezed-vacuum reservoir and the system, i.e., $\phi_{e}=\theta$ ($\theta=\pi$). When this condition is satisfied, the effective thermal noise and the two-phonon correlation strength can be reduced as
\begin{subequations}
\begin{align}
N_{s}&=\sinh^{2}(R),\\
M_{s}&=\cosh(R)\sinh(R),
\end{align}
\end{subequations}
with $R=r_n-r_e$, respectively. When $R=0$, i.e., the phonon mode is coupled to a squeezed-vacuum reservoir with $r_e=r_n$, one can find that the effective thermal noise and the two-phonon correlation strength can be suppressed completely if the phonon mode couples to a squeezed-vacuum reservoir with a specific phase and amplitude. That is to say, the squeezed-vacuum reservoir becomes an effective vacuum reservoir in our system. If $R=r_n$, i.e., the phonon mode is coupled to a vacuum reservoir, $N_s$ and $M_s$ will not equal to zero, instead, the quadratic optomechanical interaction will amplify them significantly because $r_n$ is proportional to the photon number $n$. Thus, a squeezed-vacuum reservoir with appropriate phase and amplitude is necessary for suppressing the thermal noise in our system. With the enhanced coupling strength (see Eq.~(\ref{eq:5})) as well as the completely suppressed thermal noise, our system will have obvious advantages compared with the traditional DM, as shown in the following sections.

\subsection{Squeezing of the phonon mode}

\begin{figure}[t]
\centering
\includegraphics[width=1\linewidth]{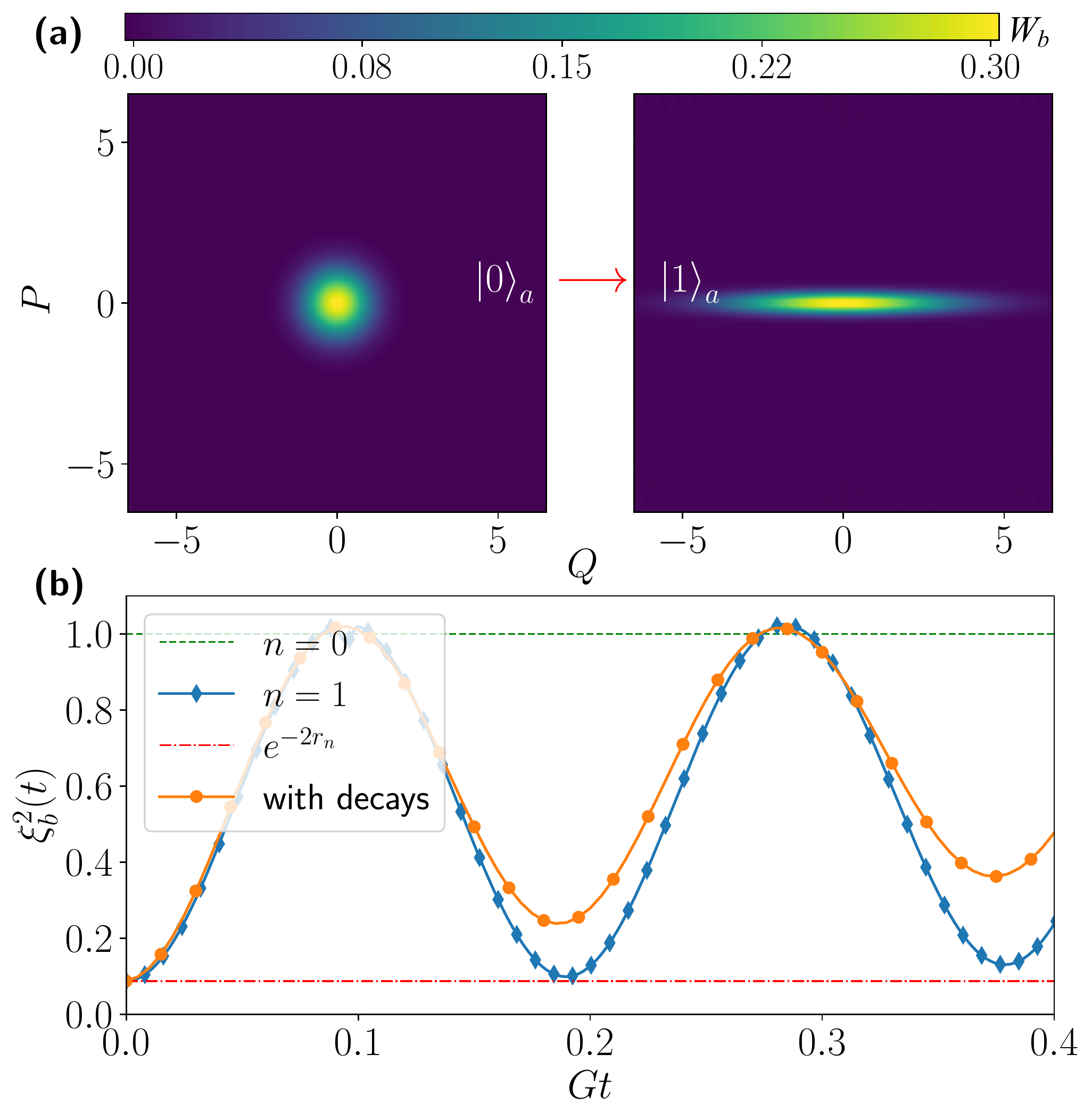}
\caption{(Color online) (a) Wigner function $W_b$ of the phonon mode when there is zero and one photon, where the quadrature variables are defined as $P=-i(b-b^\dagger)$ and $Q=(b+b^\dagger)$. (b) Evolution of the squeezing parameter $\xi_b^2(t)$ of the phonon mode with time $t$. The parameters we take for the spin ensemble are spin number $N=100$, the transition frequency $\Omega=200G$; for the phonon mode, the resonance frequency $\omega_b=2300G$; for the quadratic optomechanical coupling strength $g=0.2481\omega_b$; and for the system with decays, $\gamma=G$ and $\kappa=0.01G$.}
\label{fig:2}
\end{figure}

We first focus on the squeezing properties of the phonon mode under the quadratic optomechanical coupling. From Eq.~(\ref{eq:7}), one can find that when the photon number is not zero, the squeeze operator $S(\zeta)\neq\textbf{1}$. Obviously, we can make the phonon mode in a squeezed state by manipulating the photon number. As shown in Fig.~\ref{fig:2}(a), we numerically simulate the Wigner function $W_b$ of the phonon mode at $t=0$ when there is zero $\left(\left|0\right\rangle_a\right)$ and one $\left(\left|1\right\rangle_a\right)$ photon. We can visually see that the phonon mode is squeezed on the momentum $P$ at the expense of expanding in its amplitude $Q$ direction when $\left|0\right\rangle_a \rightarrow \left|1\right\rangle_a$. Moreover, according to the squeezing parameter of the phonon mode in Eq.~(\ref{eq:11}), we plot the evolution of the squeezing parameter $\xi_b^2(t)$ of the phonon mode with time $t$ in closed and open systems, as shown in Fig.~\ref{fig:2}(b). We can intuitively see
that when there is zero photon number, the squeezing parameter $\xi_b^2(t)=1$, and when there is a single photon, the squeezing parameter will oscillate back and forth with an approximate periodicity over time. Furthermore, one can find that the minimum of the squeezing parameter $\xi_b^2(t)$ is just the variance of the momentum $P$ at $t=0$, i.e., $\left.(\Delta P)^2\right|_{t=0}=\exp(-2r_n)$.

\subsection{Squeezing and entanglement of the spin ensemble}

As discussed above, we can get a squeezed phonon mode in our effective DM via manipulating the photon number. In this section, we will analyze the evolution of squeezing for the spin ensemble in our effective DM with system decays and study the relationship between spin-spin entanglement and spin squeezing.

\begin{figure*}
	\centering
	\includegraphics[width=1\linewidth]{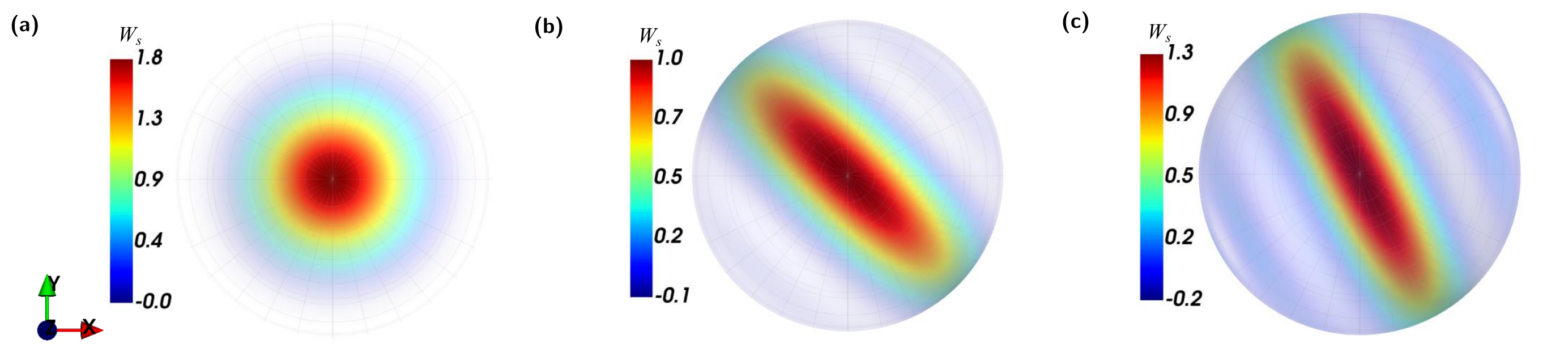}
	\includegraphics[width=1\linewidth]{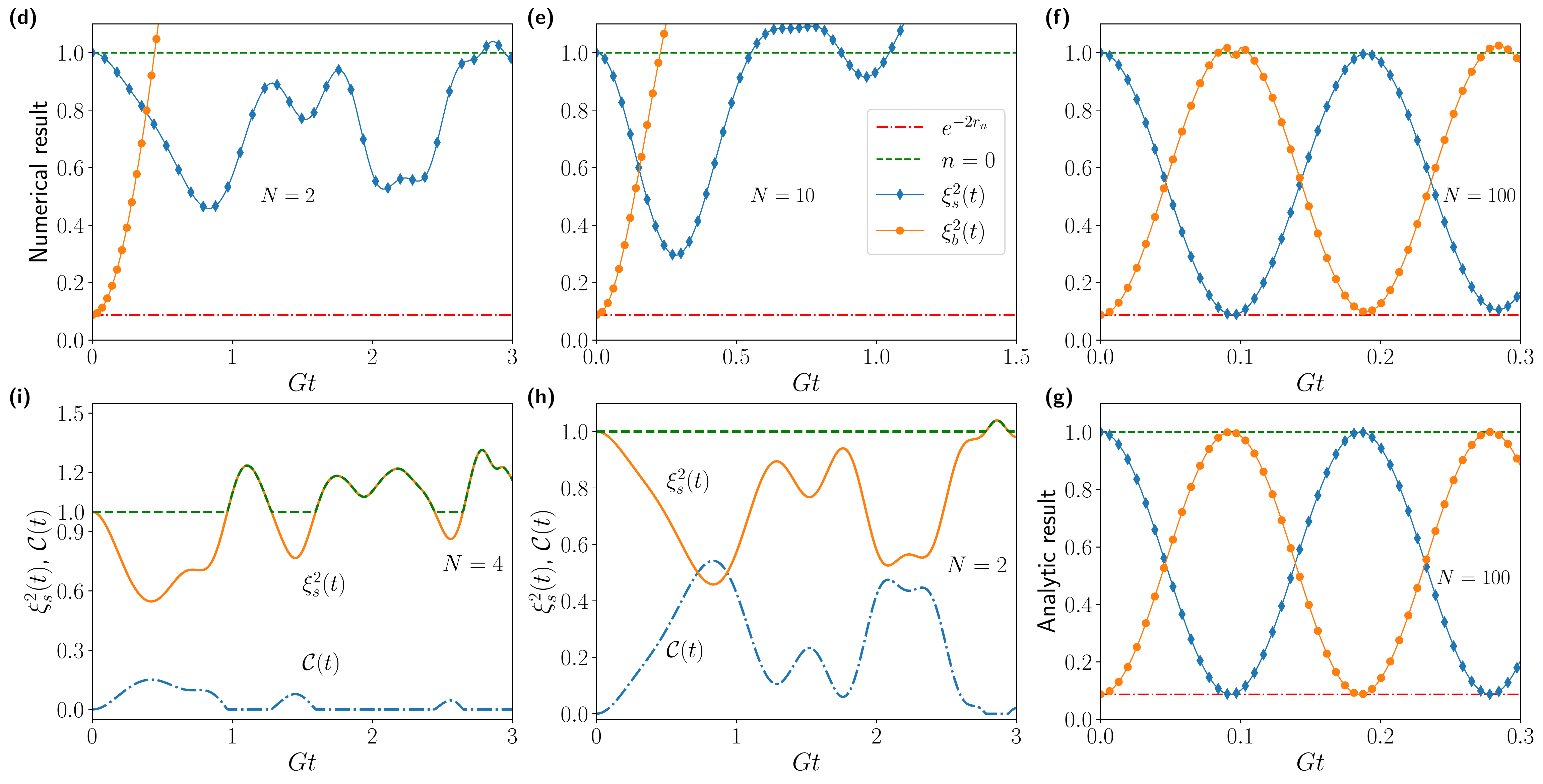}
	\caption{(Color online) (a)-(c): Evolutions of the spin Wigner function $W_s$ on Bloch sphere with spin number $N=10$ and a single photon when $Gt=0$ (a), $0.2$ (b), and $0.28$ (c). (d)-(g): Evolutions of squeezing for the spin ensemble and the phonon mode with different spin numbers ($N=2,~10,~100$) and without considering system decays, where (d)-(f) are the numerical results and (g) is the analytic result. (h) and (i): Evolutions of the spin squeezing and the spin-spin entanglement (i.e., concurrence) with different spin numbers ($N=2,~4$), where the dashed green curves represent the quantity ``$\xi_{s}^{2}+(N-1)\mathcal{C}$''. Other parameters are the same as in Fig.~\ref{fig:2}.}
	\label{fig:3}
\end{figure*}
As shown in Figs.~\ref{fig:3}(a)-\ref{fig:3}(c), we numerically simulate the evolutions of the spin Wigner function $W_s$ on Bloch sphere with spin number $N=10$ and a single photon. One can vividly observe that with the evolution, the spin Wigner function changes from isotropic distribution to being squeezed in one direction. Importantly, with the deepening of squeezing effect, light and dark interference fringes appear symmetrically in the image, and the value of negative Wigner function appears and increases with the squeezing effect. This shows that by manipulating the photon number, the spin ensemble in our effective DM can be in a squeezed state, and the spin ensemble is also in a nonclassical state. To understand this phenomenon more quantitatively, we numerically simulate the evolutions of squeezing for the spin ensemble and the phonon mode with different spin numbers and without considering system decays, as shown in Figs.~\ref{fig:3}(d)-\ref{fig:3}(f). One can find that the value of the spin squeezing $\xi_s^2$ tends to approach the initial value of the squeezing parameter $\xi_b^2$ of the phonon mode with the increase of the spin numbers, and when the spin number $N\approx 100$, squeezing transfer between the spin ensemble and the phonon mode can be observed in our system. Obviously, we can get a squeezed phonon mode via manipulating the photon number, and its optimal squeezing degree can be transferred to the spin ensemble totally when the spin number is large enough.

From the above discussion, we can make the spin ensemble in a squeezed state in our system by manipulating the photon number. Meanwhile the value of negative Wigner function implies that the spin ensemble is also in a nonclassical state. In fact, the relationship between the spin squeezing and the entanglement of spins has been deeply revealed in Ref.~\cite{XWPRA2003}. From Ref.~\cite{XWPRA2003}, we know that if the squeezing parameter $\xi_{s}^{2}\leq1$, there is a quantitative relation between the squeezing parameter $\xi_{s}^{2}$ and the concurrence $\mathcal{C}$, i.e.,
\begin{equation}\label{eq:15}
\xi_{s}^{2}=1-(N-1)\mathcal{C}.
\end{equation}
That is to say, the spin squeezing implies pairwise entanglement between spins. We will analyze whether the spin-spin entanglement can be generated in our system by manipulating the photon number and also verify the relationship in Eq.~(\ref{eq:15}).

\begin{figure}[t]
\centering
\includegraphics[width=1\linewidth]{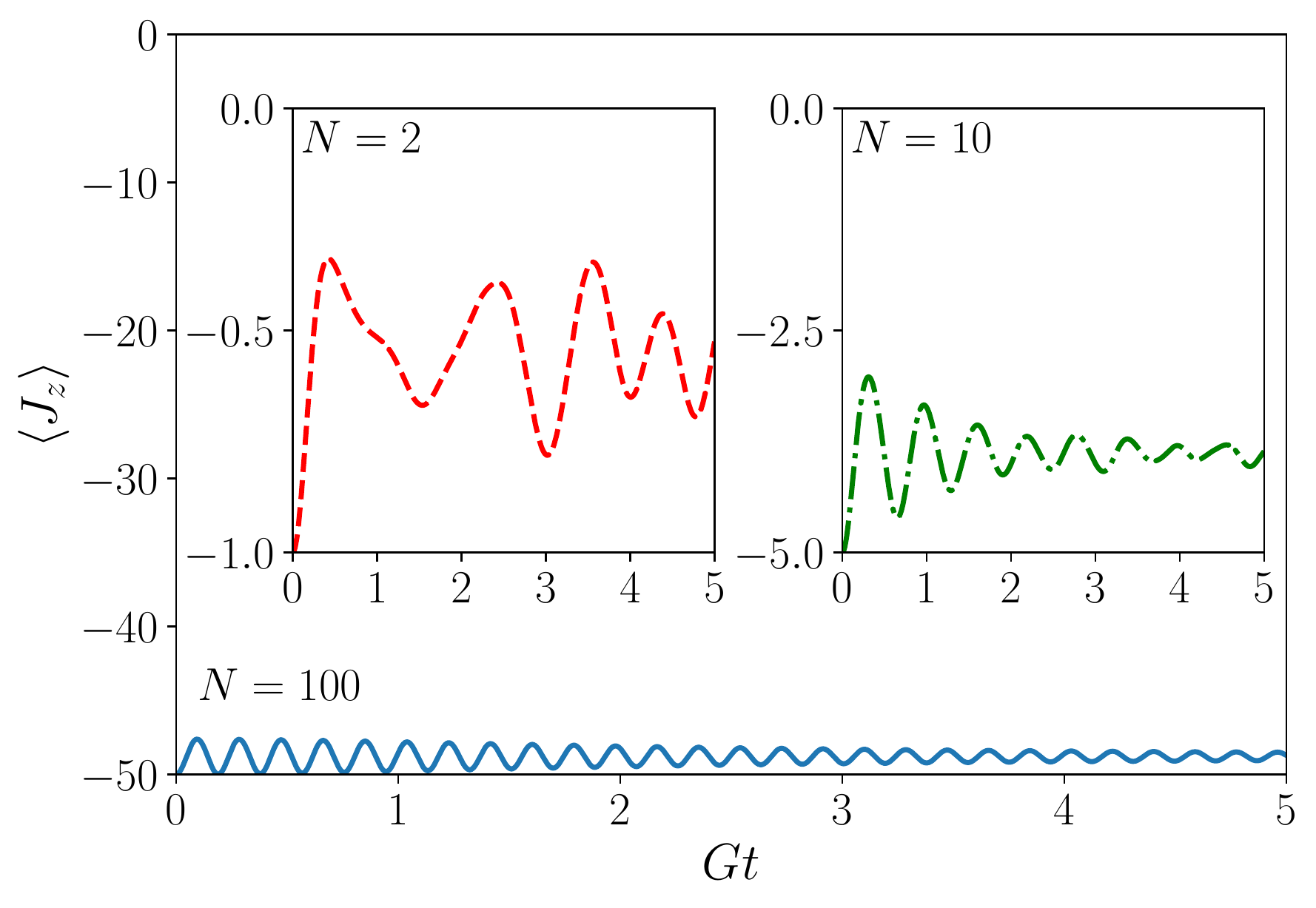}
\caption{(Color online) Evolution of the expectation value $\left\langle J_{z}\right\rangle $ of the collective operator $J_z$ with different spin numbers
($N=2,~10,~100$). Other parameters are the same as in Fig.~\ref{fig:2}.}
\label{fig:4}
\end{figure}
As shown in Figs.~\ref{fig:3}(h) and \ref{fig:3}(i), we plot the evolutions of the concurrence $\mathcal{C}(t)$ according to its original definition~\cite{WK1998}, as well as the spin squeezing $\xi_{s}^{2}(t)$ from Eq.~(\ref{eq:12}), with the spin number $N=2$ and $N=4$, respectively. One can find from the cures that pairwise entanglement of the spin ensemble can be realized in our system if and only if there is spin squeezing. Moreover, from the dashed green curves in Figs.~\ref{fig:3}(h) and \ref{fig:3}(i), which are defined as $``\xi_{s}^{2}+(N-1)\mathcal{C}"$, one can see that the quantitative relation between the squeezing parameter $\xi_{s}^{2}$ and the concurrence $\mathcal{C}$ revealed in Ref.~\cite{XWPRA2003} is fully established in our system. Please note that the concurrence $\mathcal{C} \geq 0$ in its original definition, so the dashed green curves coincide with $\xi_s^2(t)$ in the region where $\xi_s^2(t)>1$. Thus, in our effective DM, by manipulating the photon number, we can make the spin ensemble in a squeezed and entanglement state.

Physically, with the increase of the spin number $N$, the spin ensemble will remain polarized in the negative $z$ direction. As shown in Fig.~\ref{fig:4},
we plot the evolution of the expectation value $\left\langle J_{z}\right\rangle $ of the collective operator $J_z$ with different spin numbers. From the
curves, one can clearly see that the evolution of the expectation value $\left\langle J_{z}\right\rangle$ tends to be stable with time when the spin
number increases. In other words, the spin ensemble seems to be \emph{frozen} in its $-z$ direction. So it would be safe to replace
$\left\langle J_{z}\right\rangle$ with a constant, i.e, $-J$, then the squeezing parameter proposed by Kitagawa and Ueda~\cite{2} will be approximately equal to the one by Wineland et al.~\cite{3} for this special case,
i.e., $\xi_s^2\approx \xi_R^2$. In this way, we give a mathematical explanation for
the squeezing transfer between the spin ensemble and the phonon mode to better understand this interesting phenomenon. Based on the above master equation
(see Eq.~(\ref{eq:13})), we derive the differential equations of motion of the open system under the limit of a large spin number, i.e.,
\begin{equation}
\frac{d}{dt}\textbf{A}=\textbf{MA}+\textbf{$\Gamma$},
\label{eq:16}
\end{equation}
where
\begin{subequations}\begin{align}
\textbf{A}(t) &=\left(\left\langle J_{+} J_{-}\right\rangle,\!\left\langle J_{+} J_{+}\right\rangle,\!\left\langle J_{-} J_{+}\right\rangle,\!\left\langle J_{-} J_{-}\right\rangle,\!\left\langle J_{+} b\right\rangle,\right. \notag\\
&\left.\left\langle J_{+} b^{\dagger}\right\rangle,\!\left\langle J_{-} b\right\rangle,\!\left\langle J_{-} b^{\dagger}\right\rangle,\!\langle b b\rangle,\!\left\langle b^{\dagger} b\right\rangle,\!\left\langle b^{\dagger} b^{\dagger}\right\rangle\right)^{\mathrm{T}}, \\
\textbf{$\Gamma$}&=\left(0,\,0,\,0,\,0,\,0,\,0,\,0,\,0,\,\gamma M_{s}^{*},\,\gamma N_{s},\,\gamma M\right)^{\mathrm{T}}.
\end{align}
\end{subequations}
The expression of the matrix $\textbf{M}$ and the derivation process of the differential equations of motion can be found in Appendix \ref{sec:level3}.

The results of the above differential equations are very lengthy, so we do not write them here, but one can get the corresponding analytic results of the closed system from our differential equations. Then, the squeezing parameters for the spin ensemble and the phonon mode under the limit of a large spin number can be reduced as
\begin{subequations}\label{eq:19}
\begin{align}
\xi_{s}^{2}(t)&=\cos ^{2}\left[\frac{\sqrt{N} G_{n}}{2} t\right]+\xi_{b}^{2}(0) \sin ^{2}\left[\frac{\sqrt{N} G_{n}}{2} t\right], \\
\xi_{b}^{2}(t)&=\xi_{b}^{2}(0) \cos ^{2}\left[\frac{\sqrt{N} G_{n}}{2} t\right]+\sin ^{2}\left[\frac{\sqrt{N} G_{n}}{2} t\right],
\end{align}
\end{subequations}
where $\xi_{b}^{2}(0)=\exp(-2r_n)$ is the initial value of the squeezing parameter of the phonon mode. In a system with a large spin number and
considering system decays, one can find that the numerical calculation of the master equation (i.e., Eq.~(\ref{eq:13})) will greatly consume the running memory of
computer. However, the above differential equations with its initial conditions do not have this problem. From the analytic results in Eq.~(\ref{eq:19}), one
can see that the initial squeezing value $\xi_b^2(0)$ of the phonon mode can be transferred to the spin ensemble totally when
$\sqrt{N} G_{n} t/2=m\pi/2$ with $m$ being an odd number, and at this time, there is not any squeezing in the phonon mode. In Fig.~\ref{fig:3}(g), we plot
the evolutions of squeezing for the spin ensemble and the phonon mode with the analytic results. Comparing with Fig.~\ref{fig:3}(f), one can find that the analytic results
can fit the numerical results very well even if there is a little deviation between them in the long-time evolution.

\begin{figure}[t]
	\centering
	\includegraphics[width=1\linewidth]{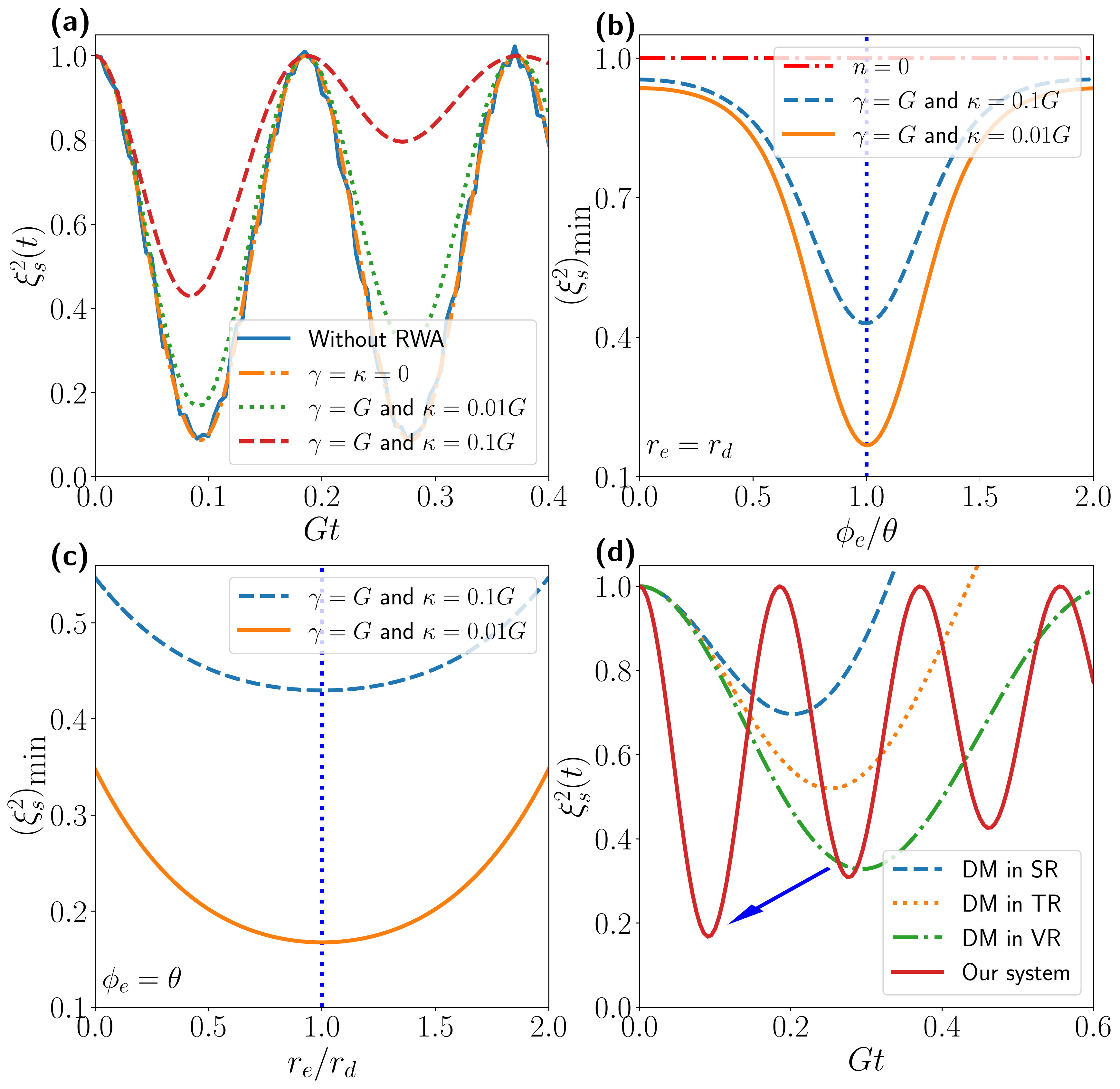}
	\caption{(Color online) (a) Evolution of the squeezing parameter $\xi_s^2$ in the system with decays; (b) and (c) minimum $(\xi_s^2)_\textrm{min}$ of the spin squeezing in the system that the parameters deviate from the the optimal squeezing phase and squeezing amplitude; (d) evolution of the spin squeezing in the open system ($\gamma=G$ and $\kappa=0.01G$), where the solid curve represents our system, besides, the dashed, dotted and dot-dashed curves represent a traditional Dicke model (DM) in a squeezed-vacuum reservoir (SR) that is the same as ours, a traditional DM in a thermal reservoir (TR) with thermal phonon number $1$, and a traditional DM in a vacuum reservoir (VR), respectively. The initial squeezing degree of the phonon mode in the traditional DM is assumed to be the same as ours, i.e., $\exp(-2r_n)$. Other parameters are the same as in Fig.~\ref{fig:2}.}
	\label{fig:5}
\end{figure}
Besides, with the differential equations (i.e., Eq.~(\ref{eq:16})), we study the evolution of the squeezing parameter $\xi_s^2$ in our effective DM with decays, as
shown in Fig.~\ref{fig:5}(a). From the curves, one can see that the periodicity of the evolution of the squeezing parameter will be destroyed in the system with decays, and the squeezing degree will gradually decrease to zero with time. However, in the short-term evolution process, the spin ensemble still has a considerable squeezing degree in our system. Moreover, we
plot the minimum of the spin squeezing that can be reached in our system with decays when the parameters deviate from the optimal squeezing phase and squeezing amplitude, as shown in Fig.~\ref{fig:5}(b) and Fig.~\ref{fig:5}(c). From the curves, one can clearly find that when the parameters deviate from the ideal parameter regime, the spin squeezing will be destroyed and more sensitive to the phase deviation. Therefore, the phase-matching condition must be satisfied for a squeezed-vacuum reservoir with a certain squeezing amplitude. This is because when the ideal parameter regime is not satisfied, the quadratic optomechanical interaction will amplify the thermal noise and destroy the spin squeezing. However, the optimal spin squeezing can be obtained in the vicinity of the ideal parameter, and importantly, it is immune to the thermal noise.

In Ref.~\cite{3}, squeezing transfer between the spin ensemble and the squeezed optical field has been confirmed in a traditional and closed DM. By squeezing transfer from the bosonic field of DM, the decoherence caused by the environment is a crucial factor for the spin squeezing. In our effective DM, the thermal noise in our system can be suppressed completely. Moreover, the interaction strength between the phonon mode and the spin ensemble is enhanced exponentially by the quadratic optomechanical coupling, i.e., $G_n=\exp(r_n)G$. These are very important for increasing and protecting the spin squeezing in an open system. As shown in Fig.~\ref{fig:5}(d), we plot the evolutions of the spin squeezing in the open environment with our system and a traditional DM, where the phonon mode in the traditional DM is assumed to have the same initial squeezing degree as our system. Besides, the traditional DM is in a squeezed-vacuum reservoir (SR) that is the same as ours, a thermal reservoir (TR) with thermal phonon number $1$, and a vacuum reservoir (VR), respectively. From the curves, one can clearly see that due to the destructive effect of the thermal noise, the squeezing degree in the traditional DM will be destroyed severely. In our system, however, due to the combined effect from the enhanced phonon-spins coupling and the suppressed thermal noise, we can obtain a better squeezing degree than the optimal squeezing that can be achieved in the traditional DM.

\section{Discussions}\label{sec:4}

In Eq.~(\ref{eq:5}) onward, the photon number operator $a^{\dagger}a$ is treated as a classical parameter rather than a dynamical variable, without considering the loss of the photon number during the cavity. From Fig.~\ref{fig:5}(a), one can find that the spin squeezing can get its optimal value when $Gt\approx 0.1$. It is worth mentioning that the collective coupling strength $G$ in the superconducting circuit can reach the order of $10\,$MHz~\cite{51}. That is to say, the photon lifetime in our system should be longer than $10^{-8}\,$s, which can be satisfied in the superconducting circuit~\cite{24}. Besides, during the derivation of the system Hamiltonian, the RWA has been used (see Eq.~(\ref{eq:6})). In Fig.~\ref{fig:5}(a), we also verify this approximation based on the differential equations of motion of the open system with the Hamiltonian without the RWA (see Appendix \ref{sec:level3} for details). One can clearly observe that the squeezing parameter without the RWA will oscillate slightly around the one with the RWA. Thus, the RWA is acceptable in the derivation of the Hamiltonian.

In our system parameters, a relatively large quadratic optomechanical coupling $g$ has been used. In principle, our theoretical scheme could be realized with a membrane-in-the-middle cavity optomechanical system~\cite{MBPRA2008} when the quadratic optomechanical coupling in this system is enhanced dramatically, and there are many
proposals to enhance the quadratic coupling, such as fiber-based cavity~\cite{57} and near-field optomechanical system~\cite{58}. To obtain a stronger quadratic optomechanical coupling, however, the good candidate for implementing our
scheme is the superconducting circuit that has the ability to simulate the quadratic optomechanical coupling~\cite{52}, as shown in Fig.~\ref{fig:1}. It has been found that the superconducting circuit can realize a stronger coupling strength compared with the one achieved in the membrane-in-the-middle cavity optomechanical system by optimizing the coupling capacitance, the bias flux and the geometrical arrangement of the circuit~\cite{52}, so that the quadratic optomechanical effect would be appreciable for a single photon in future experiments. Besides, the single-photon technology has been achieved in the superconducting circuit~\cite{59}, and other physical systems including
cold atoms~\cite{60}, diamond color centers~\cite{61} and quantum dots~\cite{62}. With the superconducting circuit and the single-photon technology, our scheme can provide an effective way to manipulate spin squeezing and entanglement and to reduce decoherence in an open system, as well as contribute to engineering new single-photon quantum devices.

\section{Conclusions}\label{sec:5}
In conclusion, we propose a single-photon-triggered spin squeezing and entanglement in a hybrid quantum model including cavity QED and optomechanics. We show that in our system, the phonon-spins coupling can be enhanced exponentially, meanwhile, the thermal noise of the phonon mode can be suppressed completely. Specifically, the squeezing of the
phonon mode triggered by a single photon can be transferred to the spin ensemble totally. Moreover, we also find that pairwise entanglement of the spin ensemble in our system can be realized if and only if there is spin squeezing. This work presents an effective way to enhance squeezing degree and reduce the decoherence in an open system, which can play an important role in the field of precision measurements and the design of new single-photon quantum devices.

\begin{acknowledgments}
This work was supported by the National Natural Science Foundation of China (Grant Nos.~11935012, 11875231, 11805073, 12175075, 11947069), the National Key Research and Development Program of China (Grant Nos.~2017YFA0304202, 2017YFA0205700), the Natural Science Basic Research Plan in Shaanxi Province of China (Grant No. 2021JQ-129) and the Scientific Research Fund of Hunan Provincial Education Department (Grant No. 20C0495).
\end{acknowledgments}

\appendix

\section{\label{sec:level1}Derivation process of the effective system Hamiltonian}

Due to the photon mode is prepared into the Fock state, so the system Hamiltonian can be rewritten as
\begin{equation}\label{eq:a1}
H=\Omega J_{z}+\left(\omega_{b}-2gn\right) b^{\dagger} b+G\left(b+b^{\dagger}\right) J_{x}-g n\left(b^2+b^{\dagger 2}\right),
\end{equation}
where the corresponding constant terms have been neglected. The last term can be diagonalized with a squeezing transformation that has the following relation,
\begin{equation}
S^{\dagger}\left(\zeta\right) b S\left(\zeta\right)=b \cosh \left(r_{n}\right)-b^{\dagger} e^{i\theta} \sinh \left(r_{n}\right),
\end{equation}
with the squeeze operator $S(\zeta)=\exp \left[\left(\zeta^{*} b^{2}-\zeta b^{\dagger 2}\right) / 2\right]$. The squeeze parameter
$\zeta=r_{n} e^{i \theta}$ with the squeeze angle $\theta=\pi$, and the squeeze amplitude can be solved as $r_n=(-1 / 4) \ln \left[1-4 n g / \omega_{b}\right]$ by setting the coefficients of these terms containing $bb$ and $b^{\dagger}b^{\dagger}$ to zero. Then we can derive the effective Hamiltonian, i.e., Eq.~(\ref{eq:5}).

\section{\label{sec:level2}Squeezing parameters with optimal squeezing for
the phonon mode and the spin ensemble}
For the phonon mode, the position and momentum amplitudes can be defined as
\begin{equation}\label{eq:a2}
Q \equiv b+b^{\dagger}, \quad P \equiv -i(b-b^{\dagger}).
\end{equation}
The phonon mode will be in a squeezed state when the variance in any direction on the $Q-P$ plane is less than 1.
Then, the quadrature squeezing of the phonon mode can be characterized by a parameter
\begin{equation}
\xi_{b}^{2}=\min _{\phi_b \in[0,2 \pi)}\left(\Delta Q_{\phi_b}\right)^{2},
\end{equation}
with $Q_{\phi_b}=be^{-i\phi_b}+b^{\dagger}e^{i\phi_b}$. From the definition of the operators $Q$ and $P$,
the squeezing parameter with optimal squeezing for the phonon mode can be derived as~\cite{4}
\begin{equation}
\xi_{b}^{2}=1+2\left(\left\langle b^{\dagger} b\right\rangle-|\langle b\rangle|^{2}\right)-2\left|\left\langle b^{2}\right\rangle-\langle b\rangle^{2}\right|.
\end{equation}
Due to $\langle b\rangle=0$ in our system, so we can get the expression of Eq.~(\ref{eq:11}).

For the spin ensemble, due to $\left\langle J_{-}\right\rangle=\left\langle J_{+}\right\rangle=0$ in our system, so the mean spin is along
the $z$ direction. Thus, the operator $J_{\vec{n}_{\perp}}$ can be written as
\begin{equation}
J_{\phi_s}=\cos (\phi_s) J_{x}+\sin (\phi_s) J_{y},
\end{equation}
with $\phi_s \in[0,2 \pi)$. Then squeezing parameter with optimal squeezing for the spin ensemble can be derived as~\cite{XWPRA2003}
\begin{equation}
\xi_s^{2}=\frac{2}{N}\left[\left\langle J_{x}^{2}\!+\!J_{y}^{2}\right\rangle\!-\!\sqrt{\left\langle J_{x}^{2}\!-\!J_{y}^{2}\right\rangle^{2}\!+\!\left\langle\left\{J_{x},\! J_{y}\right\}\right\rangle^{2}}\right],
\end{equation}
where $\left\{J_{x}, J_{y}\right\}=J_{x}J_{y}+J_{y}J_{x}$ is the anticommutator for operators $J_{x}$ and $J_{y}$, and with $J_{\pm}=J_{x} \pm i J_{y}$, we can get the expression of Eq.~(\ref{eq:12}).

\section{\label{sec:level3}Differential equations of motion in an open system}
In the Schr$\ddot{\textrm{o}}$dinger picture, the evolution of the expectation value for an arbitrary operator $O$ can be written as
\begin{equation}
\frac{d}{dt}\left\langle O\right\rangle=\textrm{Tr}\left(\dot{\rho_s}O\right).
\end{equation}
Then based on the master equation in our system (i.e., Eq.~(\ref{eq:13})), the evolution of $\textbf{A}(t)$ can be solved after some tedious calculations, i.e.,
\begin{equation}
\frac{d}{dt}\textbf{A}=\textbf{MA}+\textbf{$\Gamma$},
\end{equation}
where
\begin{widetext}
\begin{subequations}\begin{align}
\textbf{M}&=\left(\begin{array}{ccccccccccc}
-\kappa N & 0 & 0 & 0 & -i \frac{G_{n} N}{2} & 0 & 0 & i \frac{G_{n} N}{2} & 0 & 0 & 0 \\
0 & \Lambda_{1} & 0 & 0 & 0 & i G_{n} N & 0 & 0 & 0 & 0 & 0 \\
-\kappa N & 0 & 0 & 0 & -i \frac{G_{n} N}{2} & 0 & 0 & i \frac{G_{n} N}{2} & 0 & 0 & 0 \\
0 & 0 & 0 & \Lambda_{2} & 0 & 0 & -i G_{n} N & 0 & 0 & 0 & 0 \\
-i \frac{G_{n}}{2} & 0 & 0 & 0 & \Lambda_{3} & 0 & 0 & 0 & 0 & i \frac{G_{n} N}{2} & 0 \\
0 & i \frac{G_{n}}{2} & 0 & 0 & 0 & \Lambda_{4} & 0 & 0 & 0 & 0 & i \frac{G_{n} N}{2} \\
0 & 0 & 0 & -i \frac{G_{n}}{2} & 0 & 0 & \Lambda_{5} & 0 & -i \frac{G_{n} N}{2} & 0 & 0 \\
i \frac{G_{n}}{2} & 0 & 0 & 0 & 0 & 0 & 0 & \Lambda_{6} & 0 & -i \frac{G_{n} N}{2} & 0 \\
0 & 0 & 0 & 0 & 0 & 0 & -i G_{n} & 0 & \Lambda_{7} & 0 & 0 \\
0 & 0 & 0 & 0 & i \frac{G_{n}}{2} & 0 & 0 & -i \frac{G_{n}}{2} & 0 & -\gamma & 0 \\
0 & 0 & 0 & 0 & 0 & i G_{n} & 0 & 0 & 0 & 0 & \Lambda_{8}
\end{array}\right),\\
\textbf{A}(t) &=\left(\left\langle J_{+} J_{-}\right\rangle,~\left\langle J_{+} J_{+}\right\rangle,~\left\langle J_{-} J_{+}\right\rangle,~\left\langle J_{-} J_{-}\right\rangle,~\left\langle J_{+} b\right\rangle,
\left\langle J_{+} b^{\dagger}\right\rangle,~\left\langle J_{-} b\right\rangle,~\left\langle J_{-} b^{\dagger}\right\rangle,~\langle b b\rangle,~\left\langle b^{\dagger} b\right\rangle,~\left\langle b^{\dagger} b^{\dagger}\right\rangle\right)^{\mathrm{T}}, \\
\textbf{$\Gamma$}&=\left(0,~0,~0,~0,~0,~0,~0,~0,~\gamma M_{s}^{*},~\gamma N_{s},~\gamma M_{s}\right)^{\mathrm{T}},
\end{align}
\end{subequations}
and $\Lambda_{1}=2 i \Omega-\kappa N,~\Lambda_{2}=-2 i \Omega-\kappa N, ~\Lambda_{3}=i\left(\Omega-\omega_{n}\right)-\frac{N}{2} \kappa-\frac{\gamma}{2}, ~\Lambda_{4}=i\left(\Omega+\omega_{n}\right)-\frac{N}{2} \kappa-\frac{\gamma}{2},~ \Lambda_{5}=-i\left(\Omega+\omega_{n}\right)-\frac{N}{2} \kappa-\frac{\gamma}{2},~\Lambda_{6}=-i\left(\Omega-\omega_{n}\right)-\frac{N}{2} \kappa-\frac{\gamma}{2}, ~\Lambda_{7}=-2 i \omega_{n}-\gamma,$ $\Lambda_{8}=2 i \omega_{n}-\gamma$, with the following initial condition,
\begin{equation}
\textbf{A}(0)=\left(0,~0,~N,~0,~0,~0,~0,~0,~-\cosh \left(r_{n}\right) \sinh \left(r_{n}\right),~\sinh ^{2}\left(r_{n}\right),~-\cosh \left(r_{n}\right) \sinh \left(r_{n}\right)\right)^{\mathrm{T}}.
\end{equation}

Besides, we also derive the differential equations of motion based on the master equation but using the Hamiltonian without the RWA, i.e., Eq.~(\ref{eq:5}). The difference from the above differential equations is that the expression of \textbf{M} should be changed as
\begin{equation}
\textbf{M}=\left( \begin{array}{ccccccccccc}
-\kappa N & 0 & 0 & 0 & -i \frac{G_{n} N}{2} & -i \frac{G_{n} N}{2} & i \frac{G_{n} N}{2} & i \frac{G_{n} N}{2} & 0 & 0 & 0 \\
0 & \Lambda_{1} & 0 & 0 & i G_{n} N & i G_{n} N & 0 & 0 & 0 & 0 & 0 \\
-\kappa N & 0 & 0 & 0 & -i \frac{G_{n} N}{2} & -i \frac{G_{n} N}{2} & i \frac{G_{n} N}{2} & i \frac{G_{n} N}{2} & 0 & 0 & 0 \\
0 & 0 & 0 & \Lambda_{2} & 0 & 0 & -i G_{n} N & -i G_{n} N & 0 & 0 & 0 \\
-i \frac{G_{n}}{2} & -i \frac{G_{n}}{2} & 0 & 0 & \Lambda_{3} & 0 & 0 & 0 & i \frac{G_{n} N}{2} & i \frac{G_{n} N}{2} & 0 \\
0 & i \frac{G_{n}}{2} & i \frac{G_{n}}{2} & 0 & 0 & \Lambda_{4} & 0 & 0 & 0 & i \frac{G_{n} N}{2} & i \frac{G_{n} N}{2} \\
0 & 0 & -i \frac{G_{n}}{2} & -i \frac{G_{n}}{2} & 0 & 0 & \Lambda_{5} & 0 & -i \frac{G_{n} N}{2} & -i \frac{G_{n} N}{2} & 0 \\
i \frac{G_{n}}{2} & 0 & 0 & i \frac{G_{n}}{2} & 0 & 0 & 0 & \Lambda_{6} & 0 & -i \frac{G_{n} N}{2} & -i \frac{G_{n} N}{2} \\
0 & 0 & 0 & 0 & -i G_{n} & 0 & -i G_{n} & 0 & \Lambda_{7} & 0 & 0 \\
0 & 0 & 0 & 0 & i \frac{G_{n}}{2} & -i \frac{G_{n}}{2} & i \frac{G_{n}}{2} & -i \frac{G_{n}}{2} & 0 & -\gamma & 0 \\
0 & 0 & 0 & 0 & 0 & i G_{n} & 0 & i G_{n} & 0 & 0 & \Lambda_{8}
\end{array}\right).
\label{eq:a12}
\end{equation}
Then with the above differential equations of motion, we can obtain the evolution of the spin squeezing without considering the RWA, as shown in Fig.~\ref{fig:5}(a).
\end{widetext}

\end{document}